\documentclass{aa}

\usepackage{graphicx}
\usepackage{txfonts}
\usepackage[colorinlistoftodos]{todonotes}
\usepackage{color}

\usepackage{amsmath}
\usepackage{aas_macros}
\usepackage{natbib}
  \bibpunct{(}{)}{;}{a}{}{,}

\newcommand{\kms}{\mathrm{km\,s^{-1}}}

\newcommand{\cmsq}{\mathrm{cm^{-2}}}

\newcommand{\kel}{\mathrm{K}}
\newcommand{\vlsr}{v_{\rm{LSR}}}

\begin{document} 

\title{Far-infrared excess emission as a tracer of disk-halo interaction}
\subtitle{}

\author{D. Lenz \inst{1,\thanks{\email{dlenz@astro.uni-bonn.de}}}, J. Kerp \inst{1}, L. Fl\"{o}er \inst{1}, B. Winkel \inst{2}, F. Boulanger \inst{3},  G. Lagache \inst{3}}
\institute{
Argelander-Institut f\"ur Astronomie (AIfA), Universit\"at Bonn, Auf dem H\"ugel 71, 53121 Bonn, Germany
\and
Max-Planck Institut f{\"u}r Radioastronomie (MPIfR), Auf dem H\"ugel 69, 53121 Bonn, Germany
\and
Institut d'Astrophysique Spatiale (IAS), Universit\'{e} Paris-Sud XI, Orsay, France
}
\authorrunning{Lenz et al.}
\date{Received July 17, 2014; accepted October 10, 2014}

\abstract
	{Given the current and past star-formation in the Milky Way in combination with the limited gas supply, the re-fuelling of the reservoir of cool gas is an important aspect of Galactic astrophysics. The infall of \ion{H}{i} halo clouds can, among other mechanisms, contribute to solving this problem.}
	{We study the intermediate-velocity cloud IVC135+54 and its spatially associated high-velocity counterpart to look for signs of a past or ongoing interaction.}
	{Using the Effelsberg-Bonn \ion{H}{i} Survey data, we investigated the interplay of gas at different velocities. In combination with far-infrared Planck and IRIS data, we extended this study to interstellar dust and used the correlation of the data sets to infer information on the dark gas.}
	{The velocity structure indicates a strong compression and deceleration of the infalling high-velocity cloud (HVC), associated with far-infrared excess emission in the intermediate-velocity cloud. This excess emission traces molecular hydrogen, confirming that IVC135+54 is one of the very few molecular halo clouds. The high dust emissivity of IVC135+54 with respect to the local gas implies that it consists of disk material and does not, unlike the HVC, have an extragalactic origin.}
	{Based on the velocity structure of the HVC and the dust content of the IVC, a physical connection between them appears to be the logical conclusion. Since this is not compatible with the distance difference between the two objects, we conclude that this particular HVC might be much closer to us than complex C. Alternatively, the indicators for an interaction are misleading and have another origin.}


\maketitle

\section{Introduction}
\label{ch:intro}

To constantly form stars, disk galaxies such as the Milky Way need to acquire gas across cosmic times to re-fuel their gaseous reservoir \citep[e.g.][]{putman2012}. The required mass flow needs to be in excess of or equal to the star formation rate (SFR) of 1.9$\,\pm\,$0.4$\,\mathrm{M}_{\odot}\,\mathrm{yr}^{-1}$ \citep{chomiuk2011}. \citet{leitner2011} argued that recycled gas from stellar mass loss can account for much of the required matter, but an infall of pristine or sub--solar gas is required to generate the observed metallicity of stars \citep[G--dwarf problem,][]{alibes2001, schoenrich2009}. This indicates that a significant gas fraction needs to come from regions beyond the Milky Way galaxy disk.

Another potential source of star formation material is recycled disk gas, mixed with halo gas via Galactic fountains \citep{ShapiroField1976, HouckBregman1990}. Recent simulations showed that this formation channel yields a sufficient mass inflow to successfully model the evolution of the Milky Way \citep{marasco2013, fraternali2013}. The models assume a mixture of rising Galactic fountain gas with halo material of low angular momentum. The results agree well with the observed kinematics of nearby spiral galaxies \citep{fraternali2008}.

Since their discovery by \citet{muller1963}, high-velocity clouds (HVCs) have been considered to be candidates for inflowing gas. However, recent studies showed that the HVC gas accretion mass flow is only of the order of 0.08$\,\mathrm{M}_{\odot}\,\mathrm{yr}^{-1}$ \citep{putman2012}. In combination with the merging of gas-rich satellite galaxies, accretion rates of 0.1$-$0.2$\,\mathrm{M}_{\odot}\,\mathrm{yr}^{-1}$ are expected, corresponding to only 10\% of the SFR \citep{sancisi2008}.

By comparing simulations and observations, \citet{peek2008} confirmed this HVC accretion rate of 0.2$\,\mathrm{M}_{\odot}\,\mathrm{yr}^{-1}$ and furthermore concluded that neutral atomic hydrogen is just a minor (30$\,\%$) constituent of HVCs, which are thought to be predominantly ionised. These conclusions are consistent with those of \citet{barger2012} who investigated the $\rm H\alpha$ content of HVC complex A and those of \citet{fox2006} who study highly ionised species such as \ion{O}{vi} and \ion{C}{iii} for various lines of sight and deduced that HVCs have a multiphase structure.

Here, we aim to investigate observationally the role of infalling sub--solar metallicity HVC gas by studying the gas--to--dust relation of high Galactic altitude clouds.

\subsection{Interaction of HVCs with the disk--halo interface}
\label{sect:hvc_interaction}

In the past decades the interaction of HVCs with their ambient medium has been studied extensively. The Gould Belt \citep[][and references therein]{comeron1994} is a prime example for such a region because it has been proposed that it was shaped by the collision of an HVC with the Galactic disk. The Smith Cloud and its possible interaction with the coronal halo medium have been studied by \citet{lockman2008}. In their review on HVCs, \citet{wakker1997} considered the evidence for and against the interaction of a number of HVCs with the disk or intermediate-velocity clouds (IVCs) at the disk--halo interface. With the exception of HVC complex M \citep{danly1993}, the confirmed distance of HVCs, determined by absorption line studies, reaches up to around 10 kiloparsec. Thus, the neutral gas is too far away to physically interact with the gas located at the disk--halo interface.

It has been proposed that a physical connection between HVC and IVC gas is detected by \ion{H}{i} 21\,cm line emission \citep{pietz1996}. So--called velocity bridges (VB) in \ion{H}{i} appear to continuously connect the HVC and IVC in \ion{H}{i} spectra in the general direction of HVC complex C. Spatially correlated with these \ion{H}{i} VBs, \citet{kerp1999} found large--scale soft X-ray enhancements in {\it ROSAT\/} \citep{voges1999} all--sky survey data. The emission measure and plasma temperature suggest a compression of coronal halo gas at the leading edge of the HVC neutral rims, leading to a volume density enhancement by an order of magnitude at a presumed altitude of 4\,kpc. This implies that HVC gas is decelerated, the ambient halo gas being compressed and finally becoming neutral in the disk--halo interface region, which is assumed to be located at an altitude between 0.8 and 1.5$\,\mathrm{kpc}$ \citep{Richteretal2001, benbekthi2012}.

Collisions of HVCs with Milky Way disk gas force the formation of density enhancements, yielding a mixture of disk and sub--solar metallicity gas. Accordingly, targets of our search for potential interactions of HVC with the Galactic disk are IVCs with sub--solar metallicities and a close positional association with HVCs.

\subsection{IVC135+54}
\label{sect:ivc135+54}

\begin{figure}
	\includegraphics{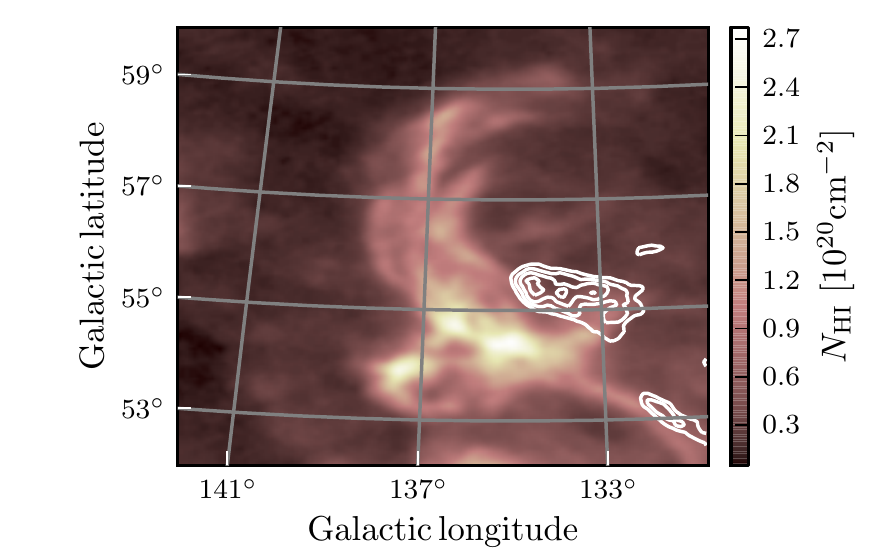}
	\caption{EBHIS \citep{kerp2011} column density map of IVC135+54 with the HVC contours superposed. For the IVC, velocities ($-52\, \kms \leq \vlsr \leq -31\, \kms$) are taken into account. For the HVC, we use ($-190\, \kms \leq \vlsr \leq -90\, \kms$). Contours start at $6\times 10^{19}\,\cmsq$ and increase in steps of $2\times 10^{19}\,\cmsq$.}
	\label{fig:ivc_mom}
\end{figure}

An example of an IVC with sub--solar metallicity and close positional correlation with an HVC is located at Galactic coordinates $(l,b, \vlsr) = (135\,^\circ, +54\,^\circ, -45\,\kms)$ (Figure \ref{fig:ivc_mom}, hereafter IVC135+54, also catalogued as IV21 in \citealt{wakker2001}). IVC135+54 has been extensively studied over the past three decades across a wide range of wavelengths. The cloud and its associated HVC are located at the north-eastern tip of the HVC complex C \citep{wakker1997}. \citet{thom2008} located large parts of complex C at a distance of approximately $10\,\mathrm{kpc}$. This distance, combined with the metallicity of $\sim 0.15\, Z_{\odot}$ \citep{fox2004}, led to the hypothesis that complex C is either accreted from the intergalactic medium (IGM), or stripped off from a satellite galaxy \citep{richter2012}.

Using \ion{Na}{i} absorption lines, \citet{benjamin1996} bracketed the distance to IVC135+54 and found a best estimate of $d = (355 \pm 95)\,\mathrm{pc}$. The high Galactic latitude of the IVC allows its height above the Galactic plane to be relatively precisely calculated at $z = (285\pm 75)\,\mathrm{pc}$. This is approximately the thickness of the \ion{H}{i} layer \citep{KalberlaKerp2009}. It is one of only seven known intermediate-velocity molecular clouds \citep[IVMCs,][]{magnani2010} that show high column densities ($\geq 10^{19}\,\cmsq$) of molecular hydrogen.

Using the full sky data of the new Effelsberg-Bonn \ion{H}{i} survey \citep[EBHIS,][]{winkel2010, kerp2011} together with Planck \citep{planck2013_vi} and IRIS \citep{miville2005} data, we investigate the interaction scenario and the implications for the transition of gas phases in the IVC. The main aim of the present work is to test whether IVC and HVC are in physical contact and to investigate the properties of dust and gas in the clouds.

The outline of this article is as follows: In Sect. \ref{ch:data} we introduce the EBHIS data of atomic hydrogen and the FIR data. In Sect. \ref{ch:analysis} we present the results from the spectroscopic \ion{H}{i} data and analyse the velocity structure of the HVC-IVC system. Furthermore, we present our Bayesian model to describe the correlation between \ion{H}{i} and dust and estimate the molecular hydrogen column density. In Sect. \ref{ch:discussion} we discuss the proposed interaction scenario and the distance discrepancy. In Sect. \ref{ch:conclusion} we summarise the results.

\section{Data}
\label{ch:data}

\begin{table*}
\centering
\caption{Properties of the data sets used in this study. $\lambda$ and $\nu$ are the reference wavelengths and frequencies. $\theta$ is the beam FWHM. $\sigma_{\rm RMS}$ is the sensitivity per beam solid angle, and $\sigma_{\rm cal}$ is the calibration uncertainty. For the spectroscopic EBHIS, $\sigma_{\rm RMS}$ is calculated for a velocity resolution of 1.45$\,\kms$.}
\begin{tabular}{c|c|c|c|c|c|c}
Survey & $\lambda$ & $\nu$ &  $\theta$ & $\sigma_{\rm RMS}$ & $\sigma_{\rm cal}$ & Reference\\
&& [GHz] & [arcmin] & & &\\
\hline
\hline
EBHIS & $21\,\mathrm{cm}$ & 1.420 & 10.8 & $\lesssim 68\,\mathrm{mJy/beam}$ & $\lesssim 3\,\%$ & 1\\
IRIS & $100\,\mathrm{\mu m}$ & 2998 & 4.30 & $0.06\, \mathrm{MJy\,sr^{-1}}$ & $13.5\,\%$ & 2\\
Planck & $350\,\mathrm{\mu m}$ & 857 & 4.63 & $0.014\, \mathrm{MJy\,sr^{-1}}$ & $\lesssim 10\,\%$ & 3\\
\end{tabular}
\tablebib{(1)~\citet{winkel2010}; (2) \citet{miville2005}; (3) \citet{planck2013_vi}}
\label{tab:data_surveyparams}
\end{table*}
We briefly present the three different data sets used for this work. For a detailed description, see the papers accompanying the data releases: \citet{kerp2011} and \citet{winkel2010} for EBHIS, \citet{miville2005} for the improved reprocessing of the IRAS Survey (IRIS) and \citet{planck2013_vi}, for example, for the high-frequency instrument onboard the Planck satellite. The parameters and properties of the data sets are summarised in Table \ref{tab:data_surveyparams}.

\subsection{EBHIS}
\label{sect:data_ebhis}

Using the Effelsberg 100-m telescope, EBHIS covers the emission of atomic hydrogen across the entire sky north of $-5^\circ$ at an effective angular resolution of $10.8\,\arcmin$. Using FPGA-FFT spectrometers \citep{klein2006} with $100\,\mathrm{MHz}$ of bandwidth, EBHIS covers not only the Milky Way, but also the local Universe out to a distance of $D \approx 280\,\mathrm{Mpc}$ or a redshift of $z\approx 0.07$. With respect to the Leiden/Argentine/Bonn Survey \citep[LAB,][]{kalberla2005}, EBHIS offers the advantage of superior resolution ($10.8\arcmin$ vs. $60.0 \arcmin$) at similar sensitivity, full angular Nyquist sampling, and coverage of both Galactic and extragalactic \ion{H}{i} emission. The average noise level is 90$\,\rm mK$ per channel ($1.29\,\rm km\,s^{-1}$) for the Galactic data. A full description of the data reduction is given in \citet{winkel2010}.

\subsection{FIR data}
\label{sect:data_fir}

Based on data of the Infrared Astronomy Satellite \citep[IRAS,][]{neugebauer1984}, \citet{miville2005} generated the improved reprocessing of the IRAS survey (IRIS) for wavelengths of $60\,\rm\mu m$ and $100\, \rm\mu m$. They used advanced algorithms and cross-calibration, mainly with the Diffuse Infrared Background Experiment (DIRBE), to improve the final data product, for instance, in terms of zero level, de-striping, calibration, and removal of zodiacal light.

The IRIS FIR data are complemented by the 2013 release of the Planck high-frequency data \citep[e.g.][]{planck2013_vi}. We used these data of Galactic dust emission to investigate the correlation of \ion{H}{i} and interstellar dust. Throughout this paper, we work with Planck data from the $857\,\rm GHz$ map because dust at typical temperatures of $20\,\rm K$ is most luminous at these frequencies. Furthermore, we used the Planck sky maps with information on the dust temperature and optical depth, extracted from a modified blackbody fit to the FIR spectrum \citep{planck2013_xi}.

\section{Analysis}
\label{ch:analysis}

To ensure a proper comparison of the different data sets and to minimise systematic uncertainties resulting from interpolations, we smooth those data with higher spatial resolution to the appropriate lower resolution of the EBHIS data by convolving them with a Gaussian kernel. The fact that the Planck beam is not a perfectly symmetric Gaussian does not have a measurable impact, as the EBHIS resolution is much lower (Table \ref{tab:data_surveyparams}).

We use data in the HEALPix format \citep{gorski2005} for pixel-based comparisons which require an equal-area projection. The presented maps are projected from HEALPix onto a  Sanson-Flamsteed grid \citep{calabretta2002} for displaying purposes.

\subsection{\ion{H}{i} data}
\label{sect:hi}

Using the EBHIS measurements of neutral atomic hydrogen, we studied the three-dimensional structure of the interstellar medium and analysed the interplay of the different velocity regimes with high spectral resolution. The Doppler radial velocity information of the \ion{H}{i} data allows us to infer the acceleration and deceleration of gas.

The EBHIS column density map of IVC135+54 is shown in Figure \ref{fig:ivc_mom}. The IVC in \ion{H}{i} consists of three distinct clumps and an extended arc-like shape that was also pointed out by \citet{hernandez2013}. In addition to the emission from IVC135+54 itself, we find \ion{H}{i} emission in the general direction of the IVC in the form of fine filamentary structures and diffuse background gas. We estimate the \ion{H}{i} mass of IVC135+54 to be 650$^{+400}_{-300}\,\mathrm{M}_{\odot}$ at a distance of 355$\,\mathrm{pc}$ \citep{benjamin1996}. The uncertainty in this mass estimate is dominated on the one hand by the distance uncertainty and on the other from diffuse and filamentary emission in the direction of IVC135+54. Thus, our calculations serve only as an order-of-magnitude estimate. Adopting this mass and spherical symmetry, we find a mean \ion{H}{i} volume density of $n = 300\,\mathrm{cm}^{-3}$.

Superposed onto the IVC in Figure \ref{fig:ivc_mom} is the HVC column density distribution in white contours. The \ion{H}{i} mass of the HVC is approximately $M_{\mathrm{HVC}} = 8\times 10^4\cdot \left(\frac{D}{10\,\mathrm{kpc}} \right)^2\,\mathrm{M}_{\odot}$.
\begin{figure}
	\includegraphics{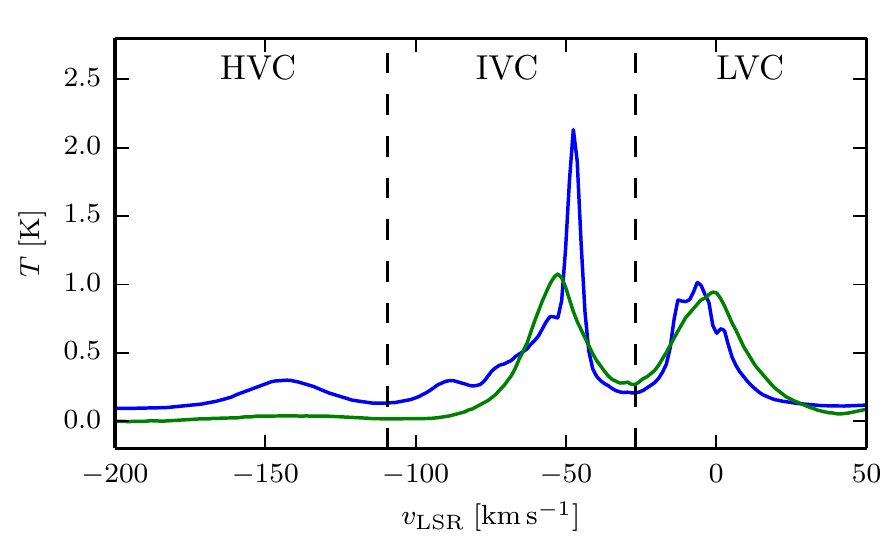}
	\caption{EBHIS median spectrum (blue) and standard deviation spectrum (green) of IVC135+54. Minima in the noise spectrum are used to separate the \ion{H}{i} components at low, intermediate, and high velocities.}
	\label{fig:noise_spectrum}
\end{figure}

\begin{figure*}
	\includegraphics[bb=0 50 530 265, clip=]{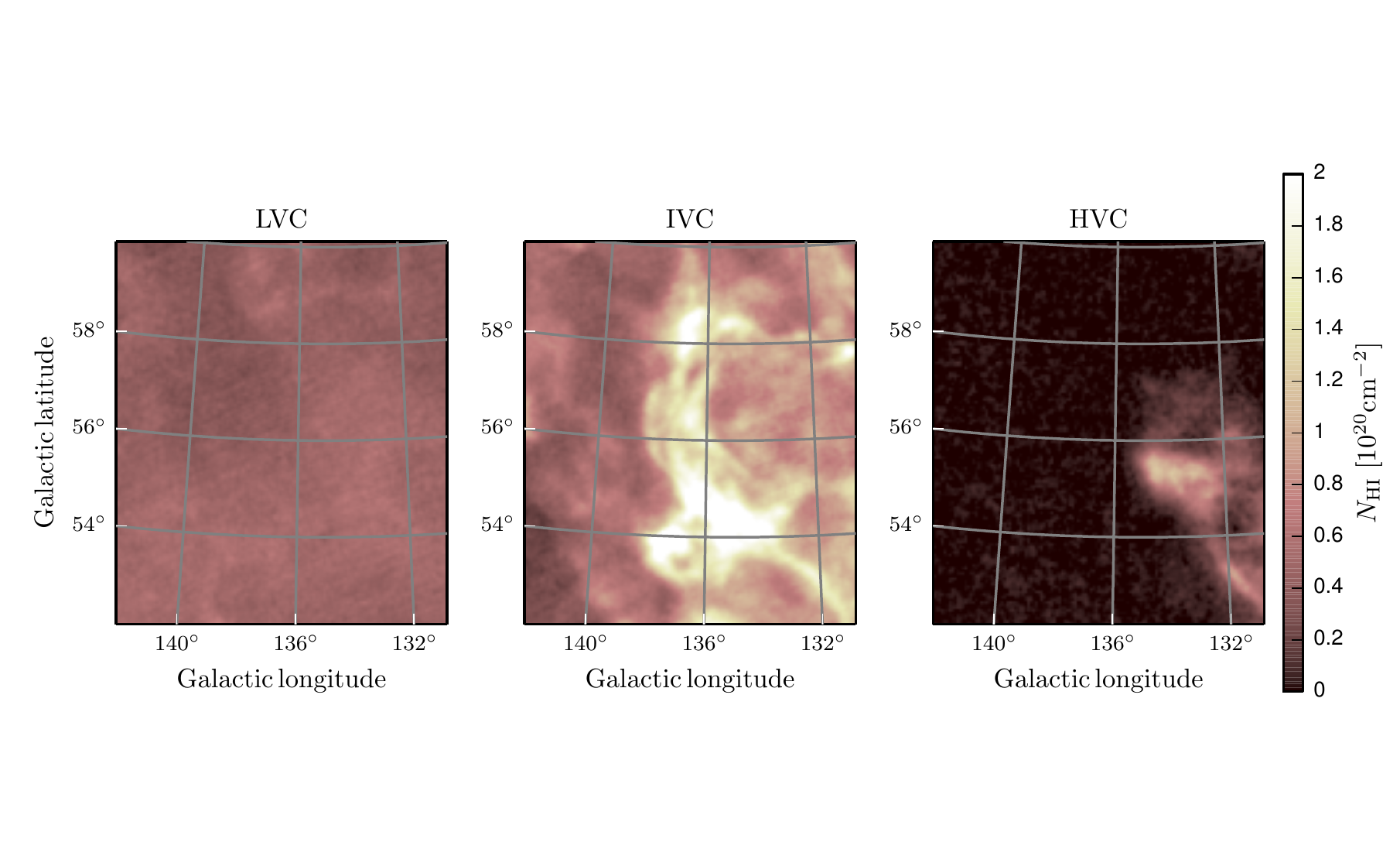}
	\caption{EBHIS column density maps of LVC, IVC, and HVC gas based on the velocity thresholds shown in Fig. \ref{fig:noise_spectrum}. The LVC component is rather faint and diffuse, while the IVC is very bright with peak brightness temperatures of 4$\,$K. For the HVC, there is hardly any background emission and the HVC has a steep brightness gradient in the eastern direction, towards the IVC.}
	\label{fig:coldenses}
\end{figure*}

An average \ion{H}{i} spectrum of the region indicated in Fig. \ref{fig:ivc_mom} is shown in Fig. \ref{fig:noise_spectrum}. For each channel of the three-dimensional datacube, we computed the median and the standard deviation. Based on the standard deviation spectrum, we decomposed the \ion{H}{i} data into three components at low, intermediate, and high velocities. Following the approach of the \citet{planck2011_xxiv}, minima in the noise spectrum were used to separate the \ion{H}{i} components. The column density maps for each component are shown in Fig. \ref{fig:coldenses}.

\subsubsection{Velocity structure}
\label{sect:hi_velocity}

\begin{figure*}
	\includegraphics{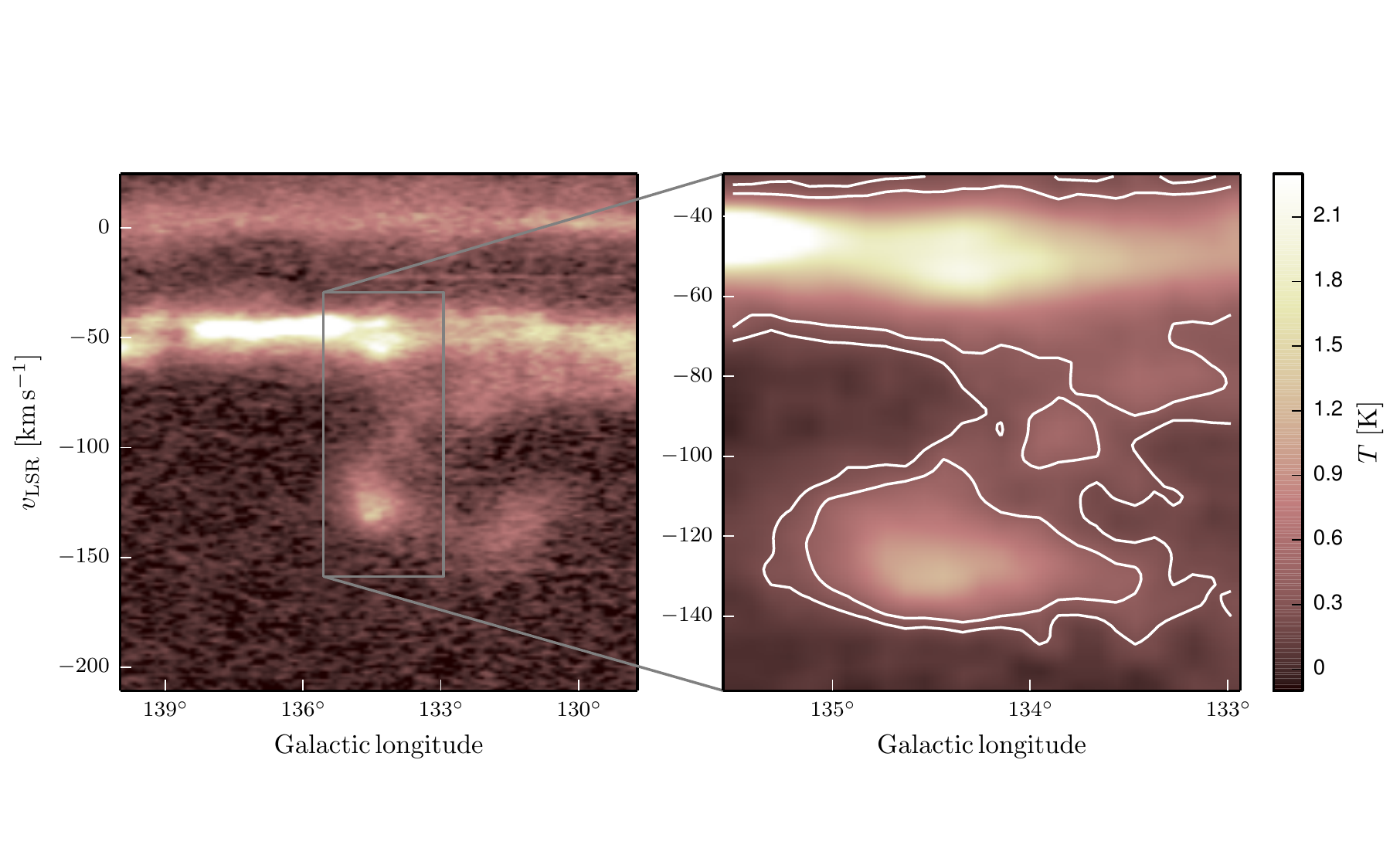}
	\caption{\textbf{Left}: EBHIS position-velocity diagram of IVC135+54 with Galactic latitude fixed at $55.8\,^\circ$. \textbf{Right}: Zoom, smoothed to a velocity resolution of $5\,\kms$. The VB connecting HVC and IVC is highlighted by white contours at the $5\sigma$-level and $8\sigma$-level, corresponding to brightness temperatures of $250\,\mathrm{mK}$ and $400\,\mathrm{mK}$.}
	\label{fig:pv}
\end{figure*}

To investigate the velocity structure and the proposed connection of high-velocity and intermediate-velocity gas, we generated a longitude-velocity diagram of IVC135+54 (Figure \ref{fig:pv}). The IVC is the brightest component, located around $\vlsr \approx -50\,\kms$. The right-hand panel is a magnified version that has been spectrally smoothed to a resolution of $5\,\kms$. Contours mark the $5\sigma$-level and $8\sigma$-level. The local gas at radial velocities around 0$\,\kms$ is rather faint with $T_{\rm B} \approx 1\,\rm K$. Towards high negative radial velocities around $-130\,\kms$, the HVC is visible. The velocity difference between the HVC components and its appearance in the column density map (Fig. \ref{fig:coldenses}) indicates a head-tail (HT) structure \citep{bruens2000}.

We confirm the velocity bridge that connects IVC and HVC gas, as detected in previous studies by \citet{pietz1996}. The VB has a brightness temperature of approximately $0.4\,\mathrm{K}$ and implies that the HVC gradually slows down and merges with the IVC in velocity space. The VB column density is in the range of $2 - 3.5\times10^{19}\,\cmsq$ and thus three times lower than the HVC, which has peak column densities around $1.2\times 10^{20}\,\cmsq$. To convert the column density into a volume density and a mass for the VB, we estimated its radial extent. Adopting the hypothesis of a physical contact between the neutral gas phase of the IVC and the HVC locates the VB at a distance of 355$\,\mathrm{pc}$ \citep{benjamin1996}. Assuming spherical symmetry, the angular extent of approximately $2\times2$ degree yields an \ion{H}{i} number density of $n_{\ion{H}{i}}\lesssim 1\,\rm cm^{-3}$ and an \ion{H}{i} mass of about 10-20$\,\rm M_{\odot}$.

\begin{figure}
	\includegraphics{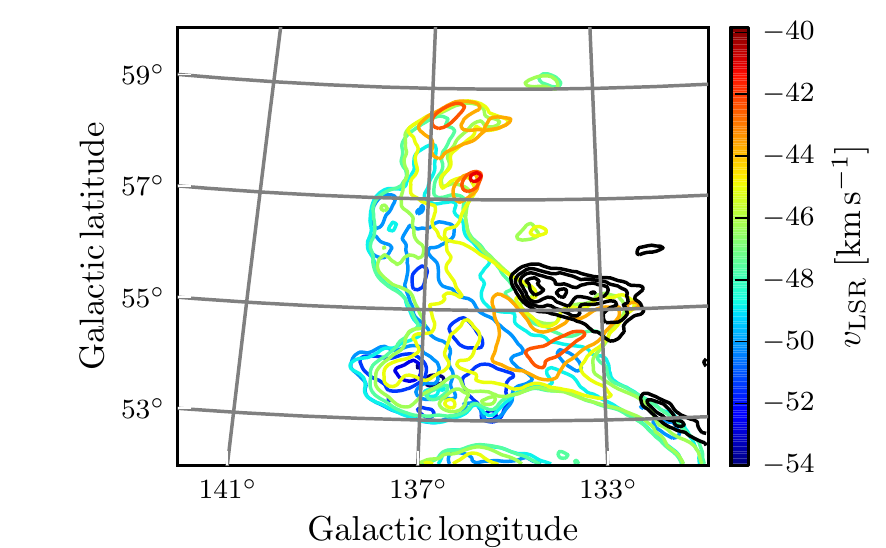}
	\caption{EBHIS Renzogram of the IVC. For each colour-coded velocity channel, a contour line at the fixed brightness temperature of $5\, \mathrm{K}$ is drawn. Superposed is the HVC with contours starting at $6\times 10^{19}\,\cmsq$ and increasing in steps of $2\times 10^{19}\,\cmsq$ (black).}
	\label{fig:renzogram_ivc}
\end{figure}

To study the IVC and its velocity structure, we generated a Renzogram \citep[e.g.][]{schiminovich1997}, which allowed us to investigate the spatial axes simultaneaously with the velocity structure (Figure \ref{fig:renzogram_ivc}). For each spectral channel of our observations ($\Delta v = 1.3\,\kms$), we drew a contour line at the fixed brightness temperature of $5\, \mathrm{K}$ and colour-coded each contour with the corresponding radial velocity. This type of plot is preferred over velocity-weighted intensity maps because it provides different information on the velocity substructure of the IVC. We also covered a broader velocity range because no averaging is involved in the Renzogram. We find a velocity shift of approximately $15\,\kms$ across the IVC. At the hypothetical impact position of the HVC, that is, between its head and the arc in the IV regime, the radial velocity shift is reduced to approximately $10\,\kms$. This is slightly different from the $15\,\kms$ shift towards the high-$b$ or low-$l$ end of the IVC and is consistent with the interpretation of momentum transfer from the HVC to the IVC.

To deduce a coherent view of IVC135+54 it is essential to consider not only the IVC and the VB, but also the HVC, which has been proposed to interact with IVC135+54. We investigate the HVC for signatures of shocks and their potential influence on the dust grains and on the hydrogen phase transitions \citep[e.g.][]{guillard2009}. The sound velocity of the HVC head is determined by fitting two Gaussian components to the HVC spectrum. The location is chosen via the HVC head column density maximum (Fig. \ref{fig:coldenses}). This returns line widths of $15.2\,\mathrm{km\,s^{-1}}$ (cold component) and $36.0\,\mathrm{km\,s^{-1}}$ (warm component). We converted the line width to upper limits of the kinetic temperature via the relation
\begin{equation}
	T_{\rm kin} \le \frac{\Delta v^2\cdot m_{\rm H}}{8k_B \ln{2}} = 21.85\cdot\left(\Delta v\,[\kms]\right)^2.
	\label{eq:tkin}
\end{equation}

The adiabatic index $\gamma$ of atomic hydrogen is 1.67. The resulting upper limits to the kinetic temperature are 8400$\,\kel$ (cold component) and $2.9\times 10^4\,\kel$ (warm component), which significantly exceeds typical temperatures (see e.g. \citet{draine2011}, who gives $\sim 100\,\kel$ and $\sim 5000\,\kel$). Hence, we suggest that the line width is dominated by turbulence and not by the kinetic temperature.

Conditions for shocks can be determined by converting the temperatures into sound velocities $c_{\rm S}$
\begin{equation}
	c_{\rm S} = \left(\frac{\gamma k_B T_{\rm kin}}{m_{\rm H}}\right)^{1/2}.
	\label{eq:cs}
\end{equation}

Based on the temperature estimates, the calculated upper limits to the sound velocities range from 6.5$\,\kms$ to 15.6$\,\kms$. Because of the large velocity difference ($\gtrsim 40\,\rm km\, s^{-1}$) between HVC and IVC, the proposed HVC impact causes supersonic shock waves.

\subsection{Dust data}
\label{sect:dust_temperature}

In addition to the \ion{H}{i} data, information on interstellar dust grains as extracted from the FIR IRIS and Planck data allow us also to probe the molecular and ionised gas phases. This analysis can only be applied to the IVC however, because the HVC is not detected in the FIR data.
For a distant HVC, the relatively weak interstellar radiation field (ISRF) at high Galactic altitudes decreases the illumination of dust grains by UV-starlight. Additionally, the suggested extragalactic origin of HVCs and especially of HVC complex C would be expected to be associated with low metallicities \citep[$\sim 0.15 \, Z_{\odot}$,][]{fox2004} and thus in small dust abundances.

\begin{figure}
	\includegraphics{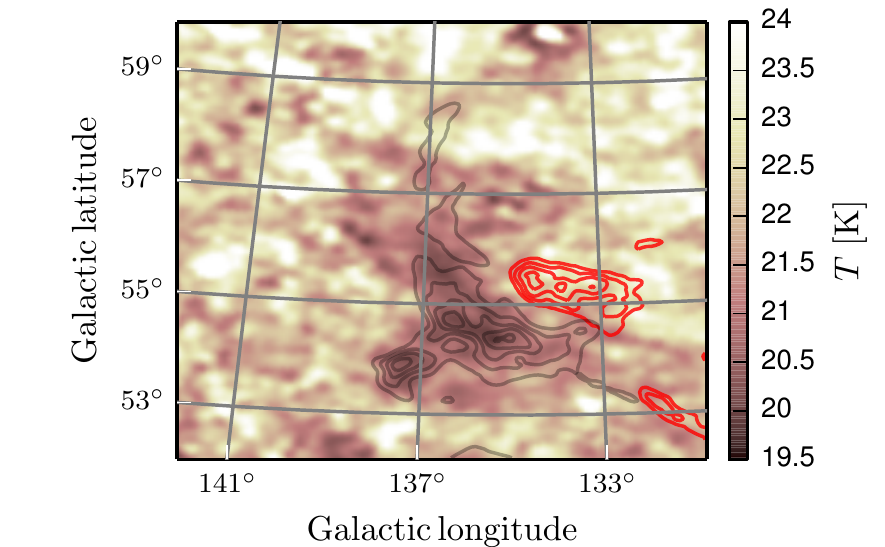}
	\caption{Dust temperature map of IVC135+54, extracted from a modified blackbody fit \citep{planck2013_xi}. Superposed are the EBHIS HVC contours, starting at $6\times 10^{19}\,\mathrm{cm^{-2}}$ and increasing in steps of $2\times 10^{19}\, \mathrm{cm^{-2}}$ (red) and the IVC contours starting at $1.1\times 10^{20}\, \mathrm{cm^{-2}}$ and increasing in steps of $4\times 10^{19}\, \mathrm{cm^{-2}}$ (grey).}
	\label{fig:dust_temperature}
\end{figure}

The dust temperature map of IVC135+54 is displayed in Fig. \ref{fig:dust_temperature} \citep{planck2013_xi}. It shows that the dust in the IVC is cooler ($\lesssim 20\,\mathrm{K}$) than in the environment ($\gtrsim 22\,\mathrm{K}$), which is associated with the higher \ion{H}{i} column densities and the presence of different molecular species in the cloud \citep[e.g.][]{weiss1999}. However, recent studies \citep{planck2013_xi} find that the coupling between dust and gas temperature is weak. Thus, dust temperature fluctuations might instead be interpreted as signature of dust evolution.

\subsection{\ion{H}{i}-dust correlation}
\label{sect:hi_dust_corr}

The correlation between neutral hydrogen and FIR dust emission has been shown to be linear up to \ion{H}{i} column densities of $1.5$ to $5\times 10^{20}\,\rm cm^{-2}$ \citep{desert1988, boulanger1996, reach1998}. For higher column densities, the FIR emission is in excess with respect to the linear model \citep{reach1998, planck2011_xxiv, planck2013_xi}.

This excess emission can partly be attributed to \ion{H}{i} self--absorption. If the assumption of optically thinness is violated, the observed \ion{H}{i} column density is systematically underestimated. However, the contribution from this process is only about a few percent in the present case because of the low \ion{H}{i} column densities in IVC135+54 \citep{planck2011_xxiv, braun2012}.

Since FIR dust emission traces not only hydrogen in the atomic phase, but actually the total hydrogen column density
\begin{equation}
	N_{\rm H} = N_{\ion{H}{i}} + 2\cdot N_{\rm H_2} + N_{\ion{H}{ii}},
	\label{eq:col_dens}
\end{equation}
the excess emission can be used to trace ionised and molecular hydrogen.

To disclose possible deviations from the correlation for instance as a result of variable dust emissivity \citep{planck2013_xi}, we used not only the FIR intensity to trace interstellar dust, but also the dust opacity
\begin{equation}
	\tau_{\nu}(\lambda) = \frac{I_{\nu}(\lambda)}{B_{\nu}(T_{\rm D})}
\end{equation}
at 353$\,\rm GHz$ provided by \citet{planck2013_xi}, which is a measure of the total hydrogen column density. We dropped the index $\nu$ because we only used this frequency.  In the following, $\tau$ is used as a measure for the dust column density for all equations and expressions related to the \ion{H}{i}-dust correlation.

To account for the different dust emissivities
\begin{equation}
    \epsilon = \tau/N_{\ion{H}{i}}
    \label{eq:emissivity}
\end{equation}
of local gas as well as intermediate-- and high--velocity gas, we adopted the approach of the \citet{planck2011_xxiv}. Hence, we interpret the total dust opacity as
\begin{equation}
    \tau(x,y) = \sum\limits_{i} \epsilon^i\cdot N_{\ion{H}{i}}^i(x,y) + Z + R(x,y).
    \label{eq:emissivities}
\end{equation}
Here, $i$ cycles over the LVC, IVC, and HVC component of the \ion{H}{i} emission presented in Fig. \ref{fig:coldenses}. $Z$ is a global offset to the full map and $R$ is the residual emission that disagrees with this linear model. To properly account for the FIR excess emission of the IVC, we modified Eq. (\ref{eq:emissivities}) to differentiate between two emissivities in the IVC regime: For low \ion{H}{i} column densities, we expect both the atomic and the ionised phase but used a single value because the FIR emissivity scales only according to the number of nucleons. At high \ion{H}{i} column densities, the molecular hydrogen contributes to the dust opacity, yielding a higher emissivity. Hence, these emissivities reflect the atomic and the molecular hydrogen of the IVC, respectively. The transition between the two emissivities $\epsilon^{\rm IVC,\,low}$ and $\epsilon^{\rm IVC,\,high}$ is realised using a step function $\theta (N_{\ion{H}{i}}^0)$. The IVC contribution to the dust opacity, $\tau^{\rm IVC}$, can then be written as
\begin{equation}
    \tau^{\rm IVC} = \epsilon^{\rm IVC,\, low}N_{\ion{H}{i}}^{\rm IVC} + \epsilon^{\rm IVC,\, high}(N_{\ion{H}{i}}^{\rm IVC} - N_{\ion{H}{i}}^{0})\cdot\theta(N_{\ion{H}{i}}^{0}).
    \label{eq:tau_ivc}
\end{equation}

\begin{figure*}
	\includegraphics[bb=0 50 530 265, clip=]{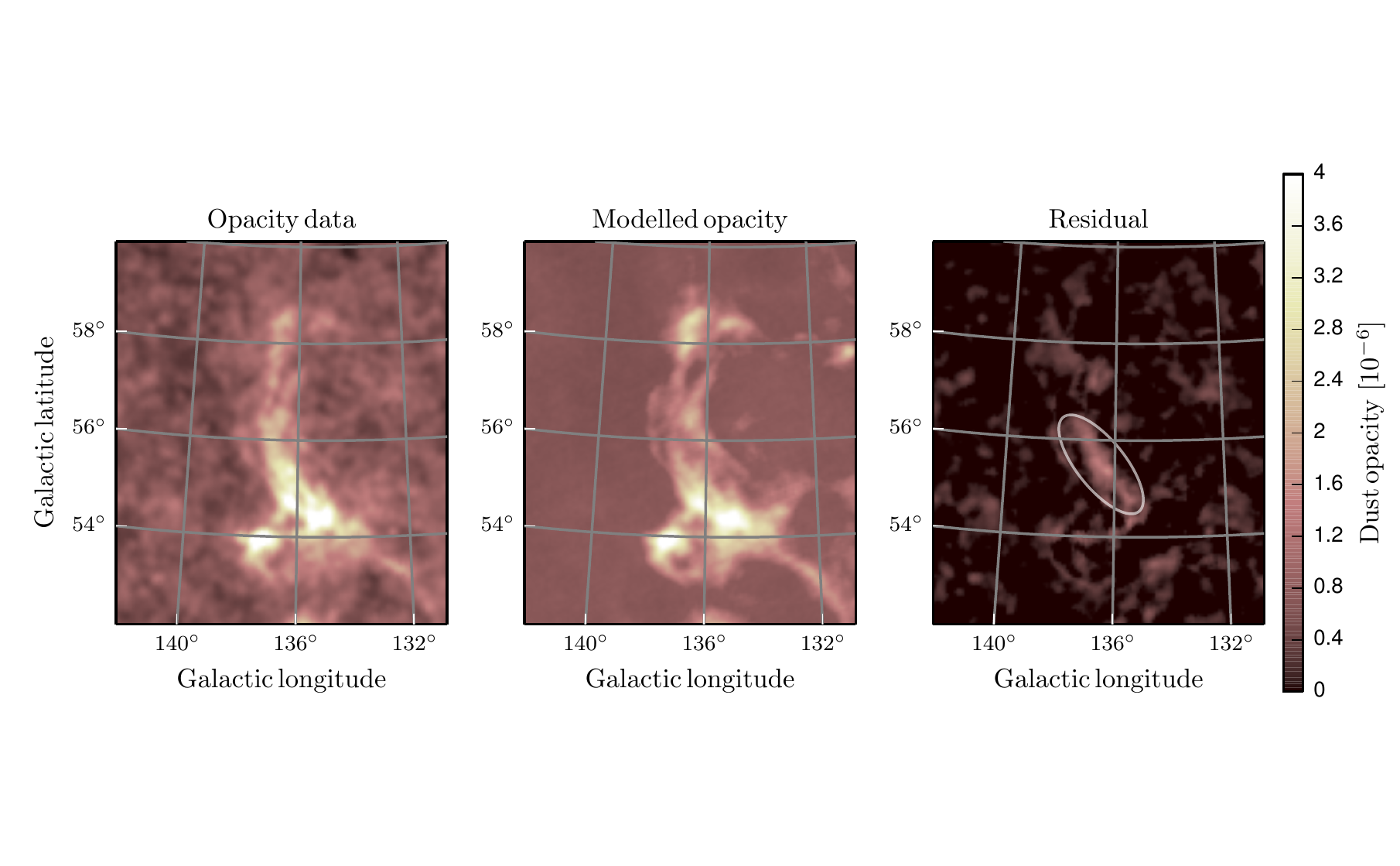}
	\caption{\textbf{Left:} IVC135+54 dust opacity at 353$\,$GHz as measured by Planck \citep{planck2011_xxiv}. \textbf{Centre:} Reconstruction of the opacity via velocity--dependent emissivities according to Eqs. (\ref{eq:emissivities}) and (\ref{eq:tau_ivc}). \textbf{Right:} Residual emission ($\tau^{\rm Data} - \tau^{\rm Model}$). The white ellipse marks the pronounced front that is connected with the HVC head.}
	\label{fig:dust_model}
\end{figure*}

\begin{figure*}
	\includegraphics[bb=0 25 600 275, clip=]{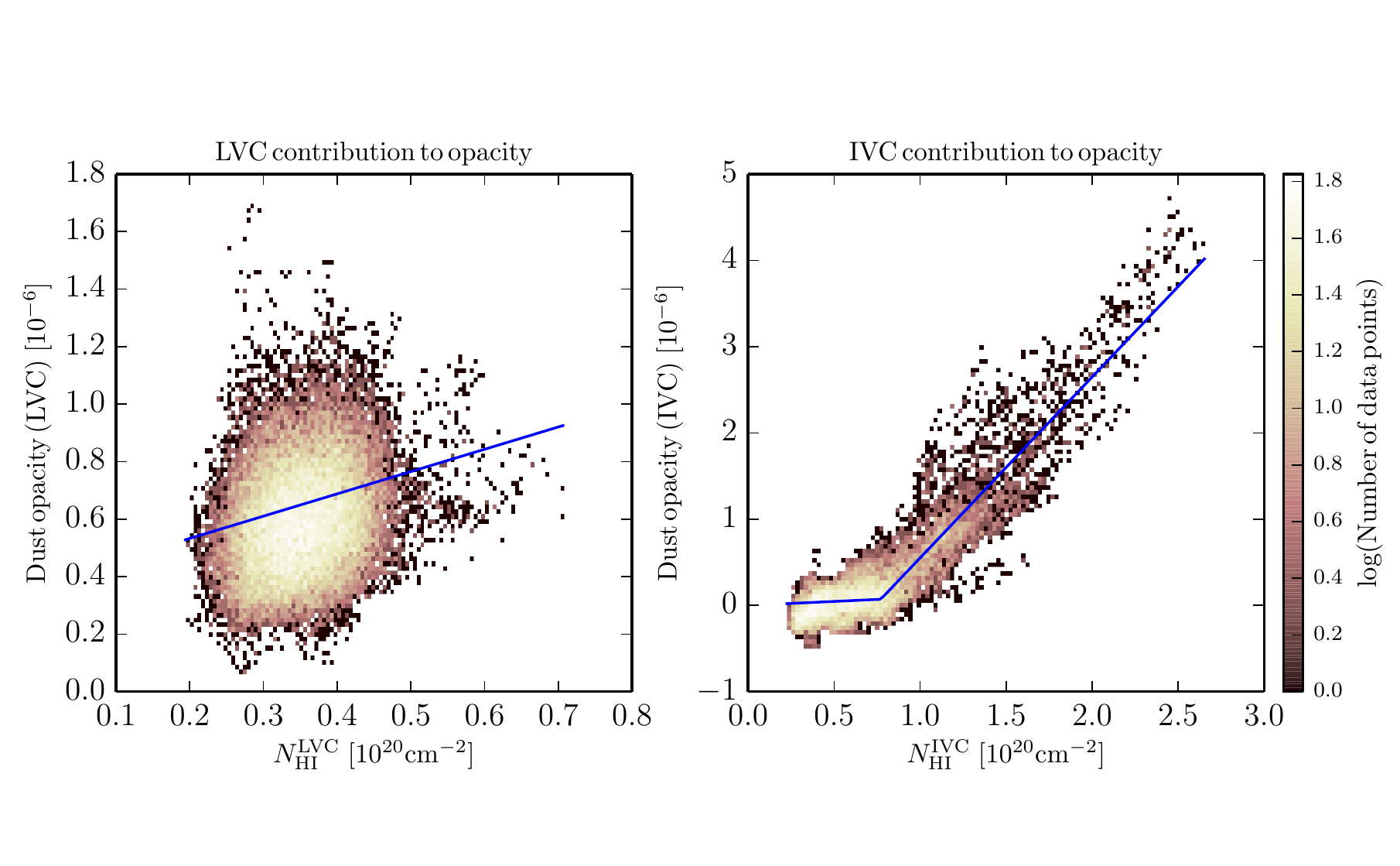}
	\caption{Dust opacity at 353$\,$GHz as a function of the EBHIS column density for the LVC and IVC gas. The \ion{H}{i} column densities correspond to the maps shown in Fig. \ref{fig:coldenses}. For the opacities, the contribution from sources other than the one shown are subtracted from the total opacity according to Eq. (\ref{eq:dust_velocity}). The linear correlation, corresponding to the emissivity $\epsilon$, is shown in blue. For the IVC, two different emissivities are used to model the correlation (Eq. \ref{eq:tau_ivc}).}
	\label{fig:histograms_opacity}
\end{figure*}

We solved Eq. (\ref{eq:emissivities}), using a Bayesian analysis realised with a Markov chain Monte Carlo (MCMC) approach that is implemented in PyMC \citep{pymc2010}. Fig. \ref{fig:dust_model} shows the result of this model.
A pixel-based evaluation of the results is shown in Fig. \ref{fig:histograms_opacity}. To compute the dust opacity for the different velocity intervals, the contribution from sources other than the one shown were subtracted from the total opacity, that is,
\begin{equation}
    \tau^{\rm i} = \tau - Z - \sum\limits_{j\neq i}\epsilon^{\rm j} N_{\ion{H}{i}}^{\rm j}.
    \label{eq:dust_velocity}
\end{equation}

The right-hand side of Fig. \ref{fig:histograms_opacity} clearly illustrates that a simple, linear approach cannot describe the data properly. Unlike the \citet{planck2011_xxiv}, for instance, we did not mask the data points that contain FIR excess emission, but included them in our model, which allowed us to better quantify this excess emission. The switchpoint $N_{\ion{H}{i}}^{0}$ of the transition is another additional parameter that is sampled by the Bayesian model. With this we also determined the column density threshold at which molecular hydrogen starts to dominate the atomic hydrogen. Table \ref{tab:emissivities} summarises the fit parameters for the \ion{H}{i} column density correlation with the FIR intensity and the dust opacity.

\begin{table}[h!]
\centering
\begin{tabular}{c|c|l}
    Data & $I_{857}$ & $\tau_{353}$\\[2pt]
    \hline
    \hline
    $\epsilon^{\rm LVC}$ & $0.44^{+0.01}_{-0.02}$ & $0.78^{+0.03}_{-0.04}$\\[3pt]
    $\epsilon^{\rm IVC,\,low}$ & $0.06^{+0.01}_{-0.01}$ & $0.09^{+0.02}_{-0.01}$\\[3pt]
    $\epsilon^{\rm IVC,\,high}$ & $1.00^{+0.00}_{-0.02}$ & $2.00^{+0.03}_{-0.02}$\\[3pt]
    $\epsilon^{\rm HVC}$ & $-0.04^{+0.02}_{-0.01}$ & $-0.09^{+0.03}_{-0.03}$\\[3pt]
    $Z$ & $0.35^{+0.00}_{-0.01}$ & $0.38^{+0.01}_{-0.01}$\\[3pt]
    $N_{\ion{H}{i}}^0$ & $0.75^{+0.01}_{-0.01}$ & $0.77^{+0.01}_{-0.01}$\\[3pt] 
\end{tabular}
\caption{Emissivities $\epsilon^{i}$ for the different velocity components, according to Eq. (\ref{eq:emissivities}). The emissivities are derived via the \ion{H}{i} column density correlation with the FIR intensity (middle column) and the dust opacity (right column). For the intensity data $I_{857}$, units are $10^{-20}\,\rm MJy\,sr^{-1}\,cm^2$ for the emissivities and $\rm MJy\,sr^{-1}$ for the offset. For the opacity data $\tau_{353}$, units are $10^{-26}\,\mathrm{cm^2}$ for the emissivities and $10^{-6}$ for the offset. For $N_{\ion{H}{i}}^0$, units are $10^{20}\,\mathrm{cm^{-2}}$. The uncertainties denote the $95\%$ uncertainty interval and only account for the statistical error of the fit. A brief investigation of the systematic uncertainties is presented in Sect. \ref{sect:displacement}.}
\label{tab:emissivities}
\end{table}

This approximation yields an emissivity $\epsilon^{\rm HVC}$ lower than zero, which cannot be motivated physically. The very low dust content of the HVC, if any, in combination with the statistical and systematical uncertainties biases the performed fit to this result. Although the MCMC error for $\epsilon^{\rm HVC}$ is small, we show below that additional systematic uncertainties have to be taken into account (Sect. \ref{sect:displacement}). One of these systematic effects is the quantitative subtraction of the FIR emission of the cosmic infrared background (CIB) that needs to be accounted for \citep{puget1996, planck2013_xi}.

As expected, the IVC emissivity at low \ion{H}{i} column densities, $\epsilon^{\rm IVC,\,low}$, is extremely low. This results from foreground opacity that is modelled via $\epsilon^{\rm LVC}$ and $Z$. Hence, $\epsilon^{\rm IVC,\,low}$ can be interpreted as systematic uncertainty to our foreground correction. For the high \ion{H}{i} column density part of the IVC, $\epsilon^{\rm IVC,\,high}$, the spatial correlation between \ion{H}{i} and dust is very tight and properly reflected by the high emissivity. This FIR excess part of the diagram allows us to infer information on the molecular hydrogen.

For the \ion{H}{i} column density $N_{\ion{H}{i}}^{0}$ at which the data deviate from a linear correlation, we find a value of $0.75\times 10^{20}\,\mathrm{cm^{-2}}$. This is lower than the commonly accepted value of $1.5$ to $5\times 10^{20}\,\mathrm{cm^{-2}}$ \citep{desert1988, boulanger1996}, but it must be noted that this transition was computed after accounting for the LVC foreground gas with median column densities of about $0.4\times 10^{20}\,\mathrm{cm^{-2}}$. Therefore, we arrive at a classical threshold of approximately $1.15\times 10^{20}\,\mathrm{cm^{-2}}$, which is consistent with previous studies. Furthermore, the pronounced front in the residual map (right-hand side of Fig. \ref{fig:dust_model}) is connected with the dense HVC head, which appears to be strongly decelerated with respect to the HVC tail.

As argued earlier, the dust opacity is more resistent to variations in the ISRF than the FIR intensity. Hence, we used the emissivities derived with the opacity for subsequent steps of the analysis. The FIR intensity-based emissivities are used for a comparison with other studies (Sect. \ref{sect:dgrs}).

\subsection{Displacement--map method}
\label{sect:displacement}

The derivation of dust emissivities $\epsilon^i$ assumes independent and identically distributed Gaussian uncertainties in the residuals ($R$ in Eq. \ref{eq:emissivities}). The right-hand side plot in Fig. \ref{fig:dust_model} clearly shows that this assumption is not valid for our data, given the structure and large--scale variations in the residual.

Therefore, we investigated the uncertainty $\sigma_{\epsilon}^{\rm pos}$, introduced by varying the position of the cloud across the sky and evaluating the model for each position shift. The displacement--map method was applied by \citet{peek2009} to investigate the dust content of different HVCs and IVCs. \citet{peek2009} aimed to obtain a reliable estimate of the uncertainties on the emissivities $\epsilon^{i}$. For a full description we refer to \citet{peek2009}, particularly to their Fig. 6.

\begin{figure*}
	\includegraphics{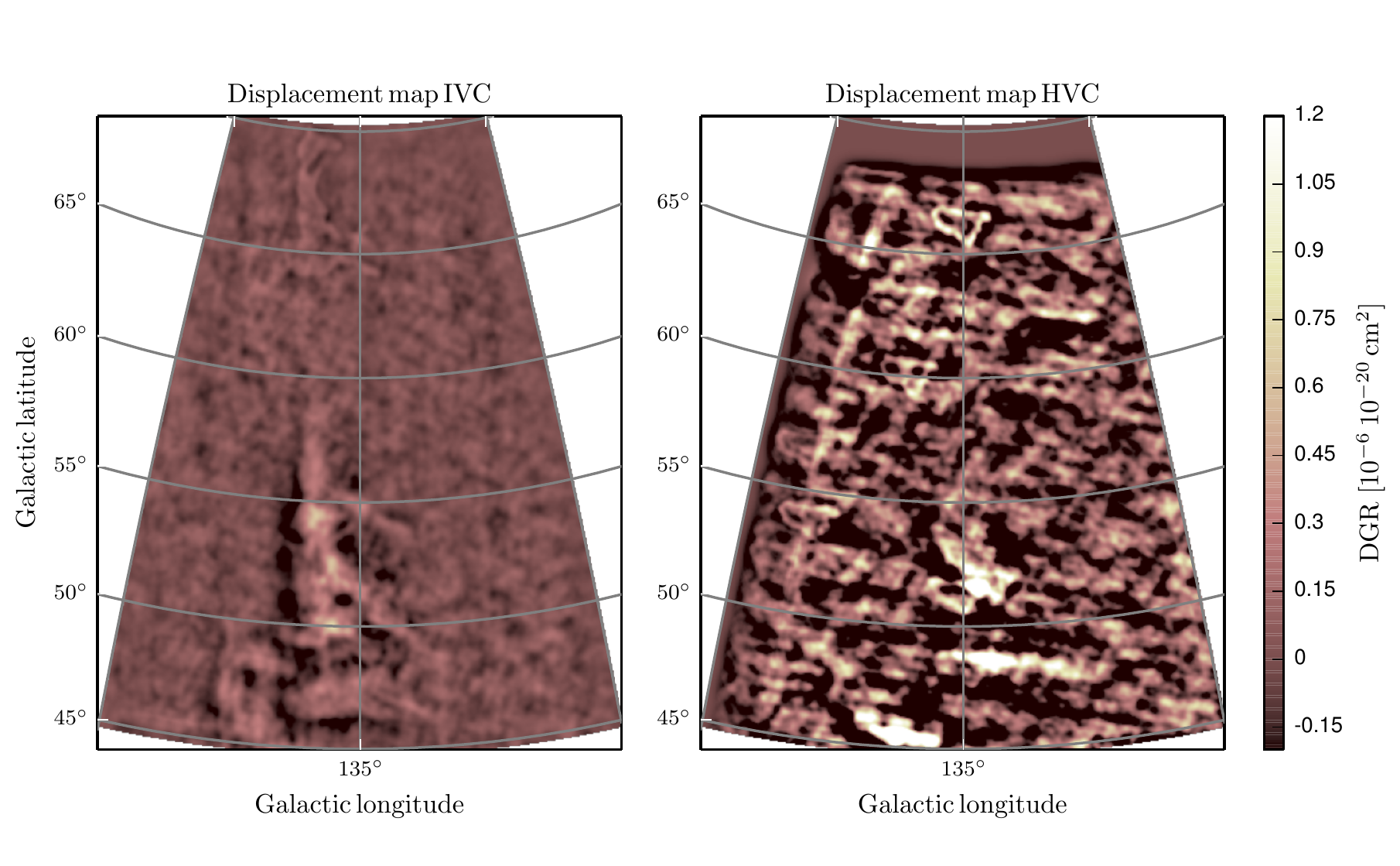}
	\caption{Application of the displacement-map technique \citep{peek2009}. \textbf{Left:} Normalised cross-correlation of IVC opacity and \ion{H}{i} column density. The variations in this map allow us to measure the uncertainty of $\epsilon^{\rm IVC}$. \textbf{Right:} Same image as left, but with the HVC displaced across the map.}
	\label{fig:displacement}
\end{figure*}

In brief, we performed template matching of the dust opacity and \ion{H}{i} column density map of IVC135+54. That is, we cross--correlated the measured \ion{H}{i} map and the calculated opacity of the IVC. This cross--correlation is highest at the location of the cloud. The amplitude of the chance correlation across the map can be used as measure of the uncertainty. The displacement--map for the IVC and HVC is shown in Fig. \ref{fig:displacement}.

To evaluate the uncertainty $\sigma_{\epsilon}^{\rm pos}$ we computed the standard deviation in signal-free regions in the displacement--map for the IVC and HVC. We find these uncertainties to be $\sigma_{\epsilon}^{\rm pos} = 0.33\times 10^{-26}\,\mathrm{cm^2}$ for the HVC and $\sigma_{\epsilon}^{\rm pos} = 0.07\times 10^{-26}\,\mathrm{cm^2}$ for the IVC. This means that the negative emissivity of $0.09\times 10^{-26}\,\mathrm{cm^2}$ for the HVC is not significant, but results from the chance correlation. For the much more luminous IVC, the displacement--map method yields an uncertainty consistent with the MCMC error estimate (see Table \ref{tab:emissivities}).

\subsection{FIR excess emission as a tracer of molecular hydrogen}
\label{sect:fir_excess}

Given the high \ion{H}{i} column densities of IVC135+54, the molecular phase is expected to dominate the ionised phase. To test this hypothesis, we used the Wisconsin H$\alpha$ mapper survey (WHAMS) data \citep{haffner2003} to estimate the column density of ionised hydrogen. Since the angular resolution of WHAMS is about  1$^{\circ}$, it does not allow a spatially precise comparison with the EBHIS and Planck data. We only evaluated the average intensity here, which yields a value of $0.46\pm 0.03\,\rm Rayleigh$ that corresponds to a column density \citep[][Eq. 2]{lagache2000} of $N_{\rm H\alpha} = 0.4\times 10^{19}\,\rm cm^{-2}$. Because this is an order of magnitude below that of neutral atomic hydrogen, neglecting the emission of ionised hydrogen is a valid first order approach. It must however be noted that with the current generation of Galactic \ion{H}{i} and dust data, we approach sensitivities at which \ion{H}{ii} can contribute to the correlation of \ion{H}{i} and dust. Accordingly, a careful treatment of the ionised phase will be mandatory for future models, given the quality of EBHIS/GASS and Planck/IRIS, to name only two.

\begin{figure}
	\includegraphics[bb=25 10 250 160, clip=]{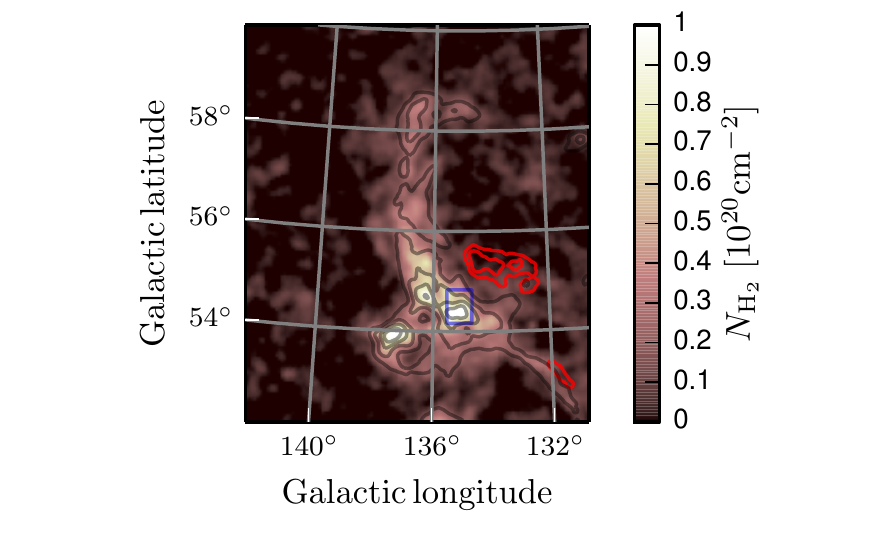}
	\caption{Distribution of the cold, dominantly molecular, gaseous phase of IVC135+54. Using Eq. (\ref{eq:molecular_column_density}), we convert the FIR excess emission (Fig. \ref{fig:dust_model} and \ref{fig:histograms_opacity}) into upper limits to the molecular column density. The blue rectangle indicates the region in which \citet{weiss1999} have detected $^{12}$CO and $^{13}$CO. Superposed are the HVC contours, starting at $6\times 10^{19}\,\mathrm{cm^{-2}}$ and increasing in steps of $2\times 10^{19}\, \mathrm{cm^{-2}}$ red) and the IVC contours starting at $1.1\times 10^{20}\, \mathrm{cm^{-2}}$ and increasing in steps of $4\times 10^{19}\, \mathrm{cm^{-2}}$ (black).}
	\label{fig:mol_coldens}
\end{figure}

Because the \ion{H}{i} column densities are not sufficiently high \citep{braun2012, planck2011_xxiv}, we do not expect signifiant \ion{H}{i} self--absorption. Therefore, the spatial information of the FIR excess emission and the emissivity $\epsilon^{\rm IVC\, high}$ describe the molecular content of the IVC. We re-arranged Eqs. (\ref{eq:col_dens}) and (\ref{eq:emissivities}) and solved for the molecular column density in the intermediate-velocity regime:
\begin{eqnarray}
    \tau^{\rm IVC} & = & \epsilon^{\rm IVC,\,high} \cdot (N_{\ion{H}{i}}^{\rm IVC,\,high} + 2\, N_{\rm H_2}^{\rm IVC})\\
    \Rightarrow N_{\rm H_2}^{\rm IVC} & = & \frac{\tau^{\rm IVC}}{2\epsilon^{\rm IVC,\,high}} - \frac{N_{\ion{H}{i}}^{\rm IVC,\,high}}{2}
	\label{eq:molecular_column_density}
\end{eqnarray}
%
%
The factor of two accounts for the fact that $\rm H_2$ consists of two hydrogen atoms. This equation can be applied to generate maps of upper limits to the molecular hydrogen column density (Fig. \ref{fig:mol_coldens}). This result must be treated carefully because it is sensitive to $\epsilon^{\rm IVC,\, high}$, which is difficult to measure properly since it requires an accurate model of the foreground LVC gas.

The information on the molecular column density can be compared with the values inferred by \citet{weiss1999} from measuring $^{12}$CO and $^{13}$CO. They mapped the small region in IVC135+54, indicated by the rectangle in Fig. \ref{fig:mol_coldens}. Using the line integral of $^{12}$CO, they found $W_{\rm CO} = 1-10\,\mathrm{K}\,\kms$. This can be converted into molecular hydrogen column densities by adopting an $\rm X_{CO}$ conversion factor of approximately $\rm X_{CO} = 4\times 10^{20}\frac{\cmsq}{\mathrm{K}\,\kms}$ \citep{draine2007b, narayanan2012}. Using this, \citet{weiss1999} deduced molecular column densities of $N_{\mathrm{H}_2} = 0.4 - 4\times 10^{20}\,\cmsq$. For the same region, we find molecular column densities in the range of $1.3\times10^{20}\,\cmsq$ to $2.1\times10^{20}\,\cmsq$, traced by the FIR excess emission (Figure \ref{fig:mol_coldens}).

Furthermore, we note that potential differences in molecular hydrogen column density derived from (a) FIR excess emission and (b) CO line--integral can arise from systematic uncertainties. In their analysis of the infrared cirrus clouds Spider and Ursa Major, \citet{barriault2010} found that the peaks of FIR excess emission and $^{12}$CO emission peaks do not coincide spatially. A follow-up analysis \citep{barriault2011} showed that low densities at the location of the FIR excess peak cause a low CO excitation temperature, which was argued to be the reason for the faint CO emission that causes the misalignment. This highlights the importance of investigating the properties of different tracers and applying them only under the proper conditions.

\section{Discussion}
\label{ch:discussion}

\subsection{Dust-to-gas ratios}
\label{sect:dgrs}

Unlike other studies of the dust-to-gas ratio (DGR) in IVCs, we did not mask the FIR excess emission, but included it in the model. Thus, a comparison with other studies is not straightforward. However, the foreground emissivities $\epsilon^{\rm LVC}$ and $\epsilon^{\rm IVC,\, low}$ can be used to compare our results with those found by the \citet{planck2011_xxiv}, who investigated the velocity-dependent emissivity of different fields at high Galactic latitudes. For the LVC and IVC component, typical emissivities are around $\epsilon_{\rm 857\,GHz} \approx 0.3-0.7\times 10^{-20}\,\rm MJy\,sr^{-1}\,cm^2$, which agrees with our results for the correlation with $I_{\rm 857\,GHz}$ (Table \ref{tab:emissivities}). Hence, IVC135+54 is an average halo cloud in terms of its DGR for low \ion{H}{i} column densities. For the DGR at higher column densities, a comparison is not possible since this part of the data is masked in the study by the \citet{planck2011_xxiv}.

For IVC135+54 in particular, the DGR can be compared with the study performed by \citet{weiss1999}, who fitted a linear correlation to the IRAS and Effelsberg data of the cloud. They did not distinguish between a linear part of the correlation and FIR excess emission. They found the mean emissivities across the entire cloud to be around $\epsilon_{\rm 3000\,GHz} \approx 1\times 10^{-20}\,\rm MJy\,sr^{-1}\,cm^2$. An inspection of the ratios of IRIS intensity at $3000\,\rm GHz$ and EBHIS total column density yields similar values between $0.7$ and $1.2\times 10^{-20}\,\rm MJy\,sr^{-1}\,cm^2$.

The latest study on IVC135+54 by \citet{hernandez2013} claims that the cloud has an unusual low DGR of $\epsilon_{\rm 3000\,GHz} \approx 0.32\times 10^{-20}\,\rm MJy\,sr^{-1}\,cm^2$, using IRAS and Green Bank Telescope data. Their values are not consistent with those reported here. The main difference probably arises from fact that \citet{hernandez2013} used IRAS instead of the reprocessed IRIS data. A direct comparison of the two data sets exhibits main differences in de-striping, calibration, and, most importantly here, on the absolute FIR luminosity level.

\subsection{Origin of the HVC--IVC system}
\label{sect:interaction}

The literature on IVC135+54 suggests different origin scenarios on the observed IVC--HVC system \citep{weiss1999, hernandez2013}. We focus on two different questions: (1) are the observed structures remnants of infalling material from the IGM, or do they result from Galactic fountains, and (2) do IVC and HVC have a common origin, or are they shaped by an HVC impact on the disk--halo interface?

\subsubsection{Galactic or extragalactic origin}

In their analysis on the \ion{H}{i}, dust, and metallicity of IVC135+54, \citet{hernandez2013} focused on question (1) and argued that the distinction between Galactic fountain and extragalactic HVC infall is only possible through evaluating the metallicity \citep[similar to][]{heitsch2009}. \citeauthor{hernandez2013} concluded that the observed sub--solar metallicity of $\log{(Z/Z_{\odot})} = -0.43\pm 0.12\,\mathrm{dex}$ is a signature of extra- or circumgalactic infall via HVCs. This proposal is consistent with the simulations of \citet{marasco2013}, showing that the scale height of the hot material of Galactic fountains is about $\sim1.2\,\mathrm{kpc}$ at $R=R_{\odot}$. Moreover, the HVC associated with IVC135+54 is part of complex C, which is proposed to be accreted from the IGM or stripped off a satellite galaxy \citep{richter2012}. This makes an extragalactic origin of the HVC most likely.

For the IVC, we find that its dust emissivity at 353$\,$GHz is similar to or higher than the emissivity of disk gas at low velocities (Table \ref{tab:emissivities}). Since the emissivity is a tracer of metallicity in first order, we conclude that the IVC has approximately solar metallicity. This agrees with the ratio of the dust intensity at 100$\,\mu\rm m$ and the \ion{H}{i} column density, which we find to be of the order of $1\times 10^{-20}\,\rm MJy\, sr^{-1}\, cm^{2}$ across the IVC. Similar ratios are found for other molecular IVCs in the Galactic halo \citep[e.g.][]{weiss1999, magnani2010}. Thus, we concluded that the IVC is of Galactic origin, whereas the HVC is associated with complex C and therefore probably of extragalactic origin.

\subsubsection{Impact onto the IVC or remnant of HVC infall}

To answer the question of whether IVC and HVC interact physically or whether both components are remnant of a larger, infalling HVC, different facts need to be taken into account. For the IVC, we found a head-tail structure (Figure \ref{fig:renzogram_ivc}), but the velocity shift of approximately $\Delta\vlsr = 14\,\kms$ can also result from a possible confusion with line-of-sight effects. \citet{bruens2000} noted that these effects can be $10\,\kms/[^{\circ}]$. This means that we do not find compelling evidence of ram-pressure interaction of the IVC with the ambient medium, as would be expected if the IVC were indeed descending through the disk--halo interface.

The assumption that the IVC-HVC system has been shaped by the interaction of a static IVC with an infalling HVC agrees with the negative spatial correlation of both clouds and also with the detected VB that connects the two components. The inferred deceleration of the HVC down to the IVC regime is a strong indicator for inelastic interaction. Furthermore, we detected an arc of FIR excess emission that indirectly traces molecular hydrogen in the IVC and is coincident with the proposed impact position of the HVC.

The FIR excess emission can originate from enhanced dust production and thus from associated formation of molecular hydrogen. We calculated the sound velocities in this regime to be 15$\,\kms$ at most, therefore the velocity difference of approximately $40\,\rm km\,s^{-1}$ between HVC and IVC implies supersonic shocks that enhance the pressure locally, which decreases the formation time of H$_2$ \citep{guillard2009, roehser2014}. The molecular hydrogen is ``invisible'' to \ion{H}{i} observations, but is consistent with the observed FIR excess emission.

\subsubsection{Distance difference between HVC and IVC}
\label{sect:distance_difference}

The distance to IVC135+54 was determined by \citet{benjamin1996} to be approximately 355$\,\rm pc$. This is 30 times closer than the $10\pm 2.5\,\rm kpc$ at which HVC complex C is located \citep{wakker2007, thom2008}. The discrepancy between the distances to these objects is inconsistent with the proposed interaction scenario. Both distances were computed via bracketing with stellar absorption lines. A detailed and helpful description of this procedure is given by \citet{schwarz1995}.

For complex C, the non-detections and thus the lower limits to the distance are only based on three sight lines that \citeauthor{thom2008} labelled S135, S139 and S441 and that are located towards the low-$l$, low-$b$ part of complex C. This part has a true angular distance to IVC135+54 of more than 30$^{\circ}$. Indeed, \citeauthor{thom2008} argued that recent data suggest that complex C extends much closer to the plane. Moreover, they noted that the low-$l$, low-$b$ part in which the sight-lines for the lower limits are located might be farther removed, thus resulting in a distance gradient across the complex.

Furthermore, the large projected angular extent of complex C of about 40$\,^{\circ}$ in combination with the assumed 10$\,\rm kpc$ distance leads to a linear extent of 8.4$\,\rm kpc$. If we drop the hypothesis that the complex is a sheet with equal distance to all points, the HVC associated with IVC135+54 can be much closer than the commonly cited 10$\,\rm kpc$, but a distance of only 350$\,\mathrm{pc}$ appears very unlikely.

To conclude the discussion on the distance discrepancy, the three-dimensional structure of complex C remains elusive, and more work is required to generate a distance map. Based on the work of \citet{thom2008}, additional sight-lines for bracketing via stellar absorption lines could improve the distance map. For the HVC related to IVC135+54, the indicators for an interaction disagree with the distance of complex C. This means that either the distance to that particular HVC is heavily overestimated, or that another, unknown process that is not a physical connection between HVC and IVC causes the velocity bridge and the associated formation of dust and molecular hydrogen in the IVC.

\section{Conclusion}
\label{ch:conclusion}

Using the latest large--scale data sets for neutral hydrogen and interstellar dust, we have investigated the well-studied cirrus cloud IVC135+54, which is located at the disk--halo interface. EBHIS, Planck, and IRIS data were used to study the gas at both intermediate and high velocities, allowing us to probe the three-dimensional structure as well as different gas phases. Given the different conclusions on the origin of the IVC--HVC system in previous studies on IVC135+54, we focussed our analysis on the origin of the clouds.

(1) We find that the high-velocity cloud has a head-tail structure in the EBHIS data, combined with a velocity bridge that continuously connects gas from the high-- to the intermediate--velocity regime. This indicates a deceleration of the high-velocity cloud.

(2) The far-infrared excess emission in IVC135+54 does not follow the distribution of either the neutral hydrogen or the dust opacity, but is instead enhanced in an arc that is connected with the high-velocity cloud head and the velocity bridge (Fig. \ref{fig:dust_model} and \ref{fig:mol_coldens}).

(3) This combination of deceleration and far-infrared excess emission gives rise to the hypothesis that the high-velocity cloud is falling onto the disk--halo interface and interacts with IVC135+54.

(4) From the FIR excess emission we created maps of molecular hydrogen that confirm the presence of $\rm H_2$ and make IVC135+54 one of the very few intermediate-velocity molecular clouds.

(5)
The overall dust-to-gas ratio of IVC135+54 is compatible with the ratios of other infrared cirrus clouds and with previous studies of IVC135+54. By comparing the low \ion{H}{i} column density part of the cloud with IRIS data at $3000\,\rm GHz$ and Planck data at $857\,\rm GHz$, we found $\epsilon_{\rm 3000\,GHz} = 0.7 - 1.2\times 10^{-20}\,\rm MJy\,sr^{-1}\,cm^2$ and $\epsilon_{\rm 857\,GHz} \approx 0.5\times 10^{-20}\,\rm MJy\,sr^{-1}\,cm^2$, respectively.

Despite the advancement in understanding IVC135+54 and its association with high-velocity cloud complex C, there are several questions that remain to be answered to complete the picture:

(a) What is the role of the ionised hydrogen? Can high-spatial resolution H$\alpha$ observations track ionisation generated by the HVC impact?

(b) The indicators for an interaction between IVC135+54 and the HVC are inconsistent with the large distances determined for other parts of HVC complex C. Could the HVC be closer to us than current distance estimations suggest, or are the indicators for an interaction misleading and have another origin?

These questions need to be addressed in future studies related to IVC135+54 to ultimately solve the question of its origin. This requires spatially extended observations of the other gas phases, that is, \ion{H}{ii} and molecular gas tracers such as CO, but also a more stringent distance determination to complex C and especially the high-velocity cloud associated with IVC135+54.


\begin{acknowledgements}
We thank the referee J. Peek for very helpful comments, especially regarding the importance of investigating the positional variation of the emissivities. We also thank I. Stewart for carefully proofreading the manuscript. We thank the Deutsche Forschungsgemeinschaft (DFG) for the project funding under grants KE757/7-1 to -3. Based on observations with the 100\,m telescope of the MPIfR (Max-Planck-Institut f\"{u}r Radioastronomie) at Effelsberg. D.L. is a member of the Bonn-Cologne Graduate School of Physics and Astronomy (BCGS). L.F. is a member of the International Max Planck Research School (IMPRS) for Astronomy and Astrophysics at the Universities of Bonn and Cologne.

\end{acknowledgements}

\bibliographystyle{aa}
\bibliography{references}


\end{document}